# Symbolic QED Pre-silicon Verification for Automotive Microcontroller Cores: Industrial Case Study

Eshan Singh[1], Keerthikumara Devarajegowda[2,3], Sebastian Simon[2], Ralf Schnieder[2], Karthik Ganesan[1], Mohammad Fadiheh[3], Dominik Stoffel[3], Wolfgang Kunz[3], Clark Barrett[1], Wolfgang Ecker[2,4], Subhasish Mitra[1]

[1]Stanford University, Stanford, CA, USA
[2]Infineon Technologies AG, Germany
[3]Technische Universität Kaiserslautern, Germany
[4]Technische Universität München, Germany

*Abstract*—We present an industrial case study that demonstrates the practicality and effectiveness of Symbolic Quick Error Detection (Symbolic QED) in detecting logic design flaws (logic bugs) during pre-silicon verification. Our study focuses on several microcontroller core designs (~1,800 flip-flops, ~70,000 logic gates) that have been extensively verified using an industrial verification flow and used for various commercial automotive products. The results of our study are as follows:
1. Symbolic QED detected all logic bugs in the designs that were detected by the industrial verification flow (which includes various flavors of simulation-based verification and formal verification).
2. Symbolic QED detected additional logic bugs that were not recorded as detected by the industrial verification flow. (These additional bugs were also perhaps detected by the industrial verification flow.)
3. Symbolic QED enables significant design productivity improvements:
(a) 8X improved (i.e., reduced) verification effort for a new design (8 person-weeks for Symbolic QED vs. 17 person-months using the industrial verification flow).
(b) 60X improved verification effort for subsequent designs (2 person-days for Symbolic QED vs. 4-7 person-months using the industrial verification flow).
(c) Quick bug detection (runtime of 20 seconds or less), together with short counterexamples (10 or fewer instructions) for quick debug, using Symbolic QED.

## 1. What is the objective of this industrial case study?

Pre-silicon verification is used to detect logic design flaws (*logic bugs*) before integrated circuits (*ICs*) are manufactured. This study focuses on logic bugs. The reasons for this focus are:

(a) Due to rapidly growing design complexity, pre-silicon verification accounts for a significant fraction of overall design effort [Foster 15]. Despite major progress, critical logic design bugs frequently escape pre-silicon verification and are detected after ICs are manufactured, during post-silicon validation or during system operation.

(b) Bug escapes that are detected during system operation in the field can have disastrous consequences, especially for critical domains such as automotive applications.

(c) During post-silicon validation, it can be extremely time-consuming and expensive to detect, localize, diagnose, and fix bugs. Hence, it is crucial that logic bugs are detected and fixed during pre-silicon verification.

Symbolic Quick Error Detection (*Symbolic QED*) [Lin 15, Singh 18] is a new approach to pre-silicon verification. Key characteristics of Symbolic QED are:

(a) It is applicable to any System-on-Chip (*SoC*) design containing at least one programmable processor core (a generally valid assumption).

(b) It is broadly applicable for logic bugs inside processor cores, accelerators, and uncore components[1].

(c) Although it is based on Bounded Model Checking (*BMC*) [Clarke 01], it does not require design-specific properties to be (manually) created.

(d) Despite use of BMC, it achieves fast runtimes and scales to billion transistor-scale designs.

(e) Results obtained using a wide variety of designs (from processor cores and accelerators to multi-core chips) demonstrate the effectiveness of Symbolic QED and its extensions for logic bugs and security threats [Fadiheh 18, 19, Ganesan 18, Lin 15, Singh 18].

We conducted this industrial case study to demonstrate the practicality and effectiveness of Symbolic QED for pre-silicon verification in an industrial setting. The specific questions that motivated our study are:

(a) What logic bugs does Symbolic QED detect (vs. those detected by state-of-the-art industrial verification flows)?

(b) How much (manual) effort is required to implement Symbolic QED in an industrial setting?

(c) How much time and compute resources are required to run Symbolic QED?

## 2. What designs were analyzed?

To analyze the effectiveness of Symbolic QED, we selected industrial designs with the following characteristics:

(a) Industrial microcontroller cores that have been thoroughly verified (over 5 years) using industrial verification techniques (especially since they target automotive applications). Thus, logic bugs detected (or not detected) by Symbolic QED can be extensively characterized.

(b) Availability of multiple designs, based on the same Instruction Set Architecture (*ISA*), to estimate the effort required in applying Symbolic QED to a new design (i.e., with an ISA for which it has not yet been used) vs. reusing Symbolic QED for subsequent designs (with the same ISA).

(c) Availability of one or more versions of each design (with RTL updates to fix bugs and add features) to analyze the effectiveness of Symbolic QED at various stages of design maturity, and also to characterize (pre-silicon) debug effort associated with Symbolic QED.

The three microcontroller core designs selected for this study, referred to as *Designs A*, *B* and *C* (Fig. 1), have been used in various automotive products. All three designs are derived from *Design 1* (Fig. 1), and they include various features targeting specific commercial products. For each design, the older RTL versions are also shown in Fig. 1 for completeness. For this case study, however, we had access to a few most recent versions of each design (Fig. 1). Hence, we have no information about the verification effort involved / bugs detected in the first *i*, *j* and *k* versions of *Designs A*, *B* and

---

[1] components in an IC that are not processor cores or co-processors, e.g., interconnect fabrics, cache / memory controllers.

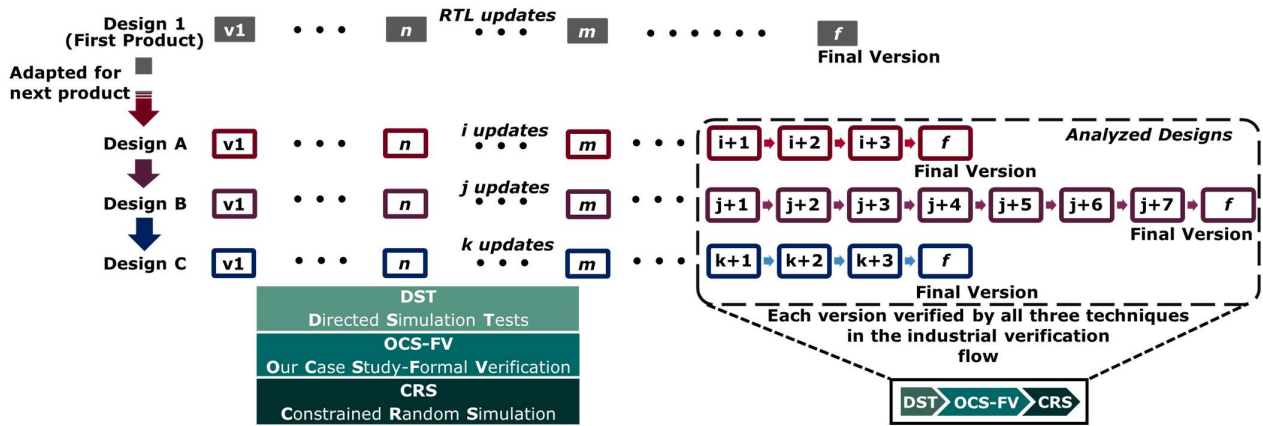

Fig. 1. Designs analyzed in this study.

C, respectively. Overall, we have 16 versions for this study: *A.v[i+1], ..., A.v[ff], B.v[j+1], ..., B.v[ff], C.v[k+1], ..., C.v[ff]*. Each version reflects an RTL update to add a feature and/or fix a bug. Not all logic bugs were fixed in RTL (e.g., some bugs were deemed to be in the design specification and were fixed there).

Each design contains approximately 1,800 flip-flops and 70,000 logic gates, and implements a custom ISA with over 50 instructions. The designs also implement built-in safety mechanisms and are classified as *ASIL-rated* (Automotive Safety Integrity Level) [ISO 26262].

There were no logic bugs reported from either post-silicon validation or field return for the final versions of each design.

### 3. How was the industrial verification flow implemented?

For each design version, the industrial verification flow used three verification techniques detailed below.

#### A. Directed Simulation Tests (DST)

During design stage, designers manually create testbenches to simulate directed test cases using commercial simulation tools. The overall objective of these tests is to verify specific features or functions. Typical checks include:

(a) Transitions between various states (e.g., reset, built-in-self-test, flush, restart, fetch, execute).

(b) Various "classes" of instructions, e.g., single- and dual-opcode instructions.

Directed tests are not meant to comprehensively verify the RTL. As a result, logic bugs escape these tests. Bugs detected by DST were fixed immediately by designers and were not recorded. Hence, those bugs aren't included in this study.

#### B. Our Case Study - Formal Verification (OCS-FV)

The verification team applied formal verification using properties generated by an in-house automation framework. We refer to this specific approach as Our Case Study – Formal Verification *(OCS-FV)*. Details about the in-house automation framework can be found in [Ecker 14]. OCS-FV creates a property for each instruction, and then proves each such property for the microcontroller core. An example property for the "ADD" instruction is shown in Fig. 2.

OCS-FV properties are different from Single-Instruction properties, introduced later in the context of Symbolic QED (explained in Question 5.C). A significant challenge in OCS-FV is to avoid false failures due to the interactions with multiple instructions. For example, the property shown in Fig. 2 does not specify if any (or which) instruction is in the execute stage at timepoint '*t*' (when the ADD instruction is decoded). As a result, the BMC tool is free to assign any (arbitrary) instruction to the execute stage at timepoint '*t*' while searching for a counterexample for the property *check_add_instruction*. Suppose that the BMC tool assigns a branch instruction to the execute stage at timepoint '*t*' and that branch causes the ADD instruction (which entered the pipeline via branch prediction) to be flushed; the property will fail – a false failure. Thus, extreme care must be taken to exclude such false failures. An additional constraint (line 3 in Fig. 2) avoids this particular case. There can be many such false failure scenarios that must be avoided by adding (manual) constraints. The resulting constraints can over-constrain the design, leading to bug escapes.

```
1 property check_add_instruction is
2 assume:
3        at t: branch_flag = '0';   // Manually added constraint
4        at t: instruction = ADD;
5        at t: core_state = decode;
6 prove:
7        at t+1: core_state = execute;
8        at t+1: result = op1 + op2;
9        at t+2: regfile_data_in = result @ t+1;
10 endproperty;
```
Fig. 2. OCS-FV property example for ADD instruction.

#### C. Constrained Random Simulation (CRS)

This verification step uses simulation testbench built following the Universal Verification Methodology (*UVM*) [UVM 17]. Designers, architects, and verification engineers first create a verification plan that includes various aspects: specific tests and methods, tools, completion criteria, resources (person, hardware, software), functions to be verified, functions not covered, etc. Verification engineers then create test cases (e.g., in SystemVerilog) based on the verification plan. Obviously, the thoroughness of such test cases depends on the verification plan. The completion criterion is often guided by code coverage and functional coverage metrics [Wile 05]. A significant challenge is to ensure the thoroughness of the verification plan and the corresponding test cases.

### 4. What effort was required to set up and run verification techniques in the industrial verification flow?

*DST Setup:* For *Design 1.v1 – 1.v[ff]* (Fig. 1), DST consumed an overall effort of 10 person-months. For each subsequent *Design A-C*, DST consumed an overall effort of 1-3 person-months for *v1* to *v[ff]*. Note that, it is difficult to isolate DST effort from general design effort since DST is applied by designers.

*DST Runtime:* For DST, a commercial simulator on an Intel Xeon E5-2690 v3 @ 2.6GHz with 32GB RAM took approximately 1 hour to simulate all directed test cases (for each design version).

*OCS-FV Setup:* For *Design 1.v1*, OCS-FV consumed 4 person-months of effort. For subsequent designs *A.v1, B.v1* and *C.v1*, it took 1 person-month effort to update existing properties. Between design versions (e.g., *A.v[2] – A.v[ff]*), the properties did not require significant updates, usually less than 1 person-hour of effort.

*OCS-FV Runtime:* For OCS-FV, Onespin 360 DV Verify (version 2016.12) on an Intel(R) Xeon E5-2690 v3 @ 2.6GHz with 32GB RAM took approximately 3 hours to run the full set of properties (for each design version).

*CRS Setup:* CRS first targeted *Design 1.v[m]*. The overall process from *Design 1.v[m]* to *Design 1.v[ff]* consumed 12 person-months of effort. The CRS was carefully designed to be extremely thorough. For subsequent designs *A-C*, the testbenches were mostly reused (with appropriate modifications). For each subsequent design *A-C*, it took 3-6 person-months of work to update the testbenches over all versions (e.g., *A.v[m]-A.v[ff]*).

*CRS Runtime:* For CRS, a commercial simulator on an Intel Xeon E5-2690 v3 @ 2.6GHz with 32GB RAM took approximately 24 hours (for each design version).

Note that, the simulation runtime to detect a bug can be shorter than running the entire set of tests (not recorded in our study).

## 5. How was Symbolic QED implemented for this study?

An overview of Symbolic QED is presented in the Appendix with emphasis on: (a) *EDDI-V* (Error Detection using Duplicated Instructions for Validation) transformation; (b) the QED module (added to the fetch unit of a processor core during BMC) which is only used by the BMC tool (i.e., it is not added to the manufactured chip). As also explained in the Appendix A.1, Symbolic QED needs to start from a *QED-consistent* architectural state to avoid false failures. We used a starting state with the core in operating mode, the pipeline empty, and all registers and memory locations equal to 0. For our designs, these conditions ensure that (in the absence of a bug) QED sequences from this state execute correctly. We used the QED module (Appendix A.2) from [Ganesan 18] (seen in Fig. 4(a)), which is an enhanced version of the QED module in [Lin 15, Singh 18]. We further enhanced EDDI-V as follows.

### A. Enhanced EDDI-V: Control flow errors

We extended EDDI-V to detect bugs that cause control flow errors (*control flow bugs*) as well. The CFCSS-V Symbolic QED property [Singh 18] detects some control flow bugs, but misses those causing incorrect branch directions. In our Enhanced EDDI-V approach targeting control flow errors, we use a QED-Control Flow *(QED-CF)* module (Fig. 5) which is instantiated between the QED module (from [Ganesan 18] [2]) and the instruction fetch (*IF*) stage of the core (Fig. 4). The QED-CF module captures the target address of each control flow instruction (branch or jump), and compares the target of an original control flow instruction with the corresponding duplicate instruction. If the targets match, BMC continues with the QED sequence. Upon mismatch, the BMC tool can inject any instruction (it chooses), to represent an incorrect control flow – this injected instruction then causes an EDDI-V QED check failure. In Fig. 3, the bug causes the duplicate sequence to incorrectly take the conditional branch BEQZ (branch if previous result is equal to zero). Since the target address of the original branch (*2*) mismatches with the duplicated branch (*5*), the BMC tool chooses to execute the load immediate instruction in address 5 (instead of the ADD instruction in address 2), causing a QED check failure (*R4* vs. *R12*).

---

[2] The designs in this study did not feature interrupts or generate stalls. As a result, the *fetch_next* signal of the QED module was constrained to 0 using the BMC tool.

| Original Sequence | Duplicate Sequence |
|---|---|
| R2 = 2, R3 = 1 | R10 = 2, R11 = 1 |
| 0: R1  = R2 - R3 | 0: R9  = R10 - R11 |
| 1: BEQZ   #5 | 1: BEQZ   #5 |
|   // R1 != 0 |   // R9 != 0 |
|   // Branch not taken | **// Error: Branch taken** |
| 2: R4  = R2  + R3 |   // Ignore QED instruction |
|   // R4 = 3 |   // 2: R12 = R10 + R11 |
|  | **// Use Non-QED instruction** |
|  | 5: LD R12 0 |
|  |   // R12 = 0 |

Fig. 3. Enhanced EDDI-V: control flow errors example.

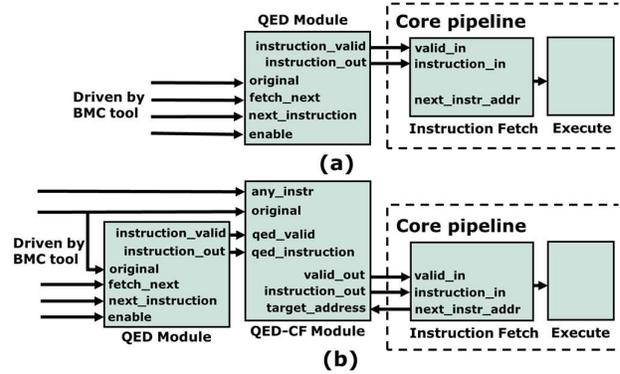

Fig. 4. The location (a) of the QED module, and (b) with the QED-CF module inserted.

To avoid false failures, our QED-CF module imposes two conditions: (a) for control flow instructions that use conditional flags (e.g., *branch if previous instruction result is zero*, as in Fig. 3), an original (duplicate) control flow instruction can only follow an original (duplicate) instruction; (b) the instruction following an original (duplicate) control flow instruction must also be an original (duplicate) instruction. The second condition ensures that if a control flow instruction causes a pipeline flush, only the next instruction from the same sequence (original or duplicate) will be flushed with it (our microcontroller core designs contain 2-stage pipelines). These conditions are sufficient for in-order, 2-stage cores, the designs analyzed in this study.

The inputs to the QED-CF module are: 1) *qed_valid*, output of the QED module (shown in Fig. 4); 2) *qed_instruction*, output of the QED module; 3) *original*, BMC signal (also an input to the QED module) to execute an original (when high) or duplicate (when low) instruction; 4) *target_address*, the next instruction address (next_instr_addr in Fig. 4) fetched by the IF stage; 5) *any_instr*, a valid, unconstrained instruction selected by the BMC tool but importantly not duplicated by the QED module. The outputs are 1) *valid_out*, the valid bit from the QED module 2) *instruction_out*, the instruction fetched by the IF stage to execute in the pipeline. The internal variables are: 1) *orig_queue*, a queue data structure that stores target address of the original branch instructions; 2) *is_CF*, determines if *qed_instruction* is a valid control flow instruction; 3) *was_CF*, is *is_CF* buffered by 1 cycle to let the branch execute; 4) *was_orig*, is *original* buffered by 1 cycle to let the branch execute; 5) *head_target*, the previous head of the queue; 6) *match*, true when the duplicate control flow instruction target matches original instruction target; 7) *orig_nCF_match*, true for original instructions, non-control flow duplicate instructions, or control flow duplicate instructions that match the original target. Depending on this value, *instruction_out* either outputs the next *qed_instruction* to the IF stage or the unconstrained instruction from the BMC tool, *any_instr*. This QED-CF module is designed for a 2-stage in-order pipeline.

```
INPUT: qed_valid, qed_instruction, original, target_address, any_instr
OUTPUT: valid_out, instruction_out

// internal variable initialization
orig_queue ← 0;
was_CF ← 0;
was_orig ← 0;
// end initialization
valid_out ← qed_valid;
is_CF ← qed_instruction[opcode_bits] = CF_opcode ? qed_valid : 0
was_CF ← is_CF;            // buffer by 1 cycle till branch executes
was_orig ← original;
if (was_orig && was_CF):     //when original branch executes
   orig_queue.push(target_address);   // store branch target in queue
else if(!was_orig && was_CF):   // when duplicate branch executes
   head_target ← orig_queue.pop();    // remove target at queue head
   match ← (orig_target = head_target) ? 1 : 0;     //check target
orig_nCF_match ← was_orig | !was_CF | match;
instruction_out ← orig_nCF_match ? qed_instruction : any_instr;
```

Fig. 5. Pseudo code for QED-CF module.

### B. Enhanced EDDI-V: Duplication using memory

Unlike the EDDI-V overview in Appendix A.1, we may choose to not divide the register space into two halves. Instead, the (original and) duplicate values may be (temporarily) stored in memory (Fig. 6) for various reasons: certain instructions may only use specific registers (e.g., *load-immediate* might only load into R0) or a bug may be triggered only when certain specific registers are used. The QED module for EDDI-V (in [Ganesan 18]) is modified accordingly to insert additional loads and stores each time it switches between original and duplicate instructions. During this switch, register values for the original instructions are first saved in (corresponding) memory locations. Next, for the duplicate instruction sequence, source registers are loaded from (corresponding) memory locations. After the duplicate instruction sequence completes, the updated registers are loaded back into memory. A QED check now compares contents of corresponding original and duplicate memory locations. To avoid unnecessary loads and stores, our updated QED module (for Enhanced EDDI-V: duplication using memory) tracks registers in use with two bits per register: a source and a destination bit. Initialized to 0, the source (destination) bit is set to 1 if the register content is used as source operand (or written to as destination). If a source bit is 1, the register doesn't need to be reloaded from memory. If a destination bit is 1, then the register shouldn't be loaded from memory and should to be written back to memory at the end of the sequence. These bits are reset (to 0) at each transition between original and duplicate instruction sequences.

```
Original Sequence              Duplicate Sequence
//Load source reg              //Load source reg
LD  R2, [R2_orig_mem]          LD  R2, [R2_dup_mem]
LD  R3, [R3_orig_mem]          LD  R3, [R3_dup_mem]
R1  = R2 + R3                  R1  = R2 + R3
LD  R5, [R5_orig_mem]          LD  R5, [R5_dup_mem]
R4  = R5 - R2                  R4  = R5 - R2
R1  = R2 + R4                  R1  = R2 + R4
//Store updated reg            //Store updated reg
ST  R1, [R1_orig_mem]          ST  R1, [R1_dup_mem]
ST  R4, [R4_orig_mem]          ST  R4, [R4_dup_mem]
```

Fig. 6. QED duplicated sequence using memory.

### C. Single-Instruction Properties (Single-I)

For each instruction in the ISA, we specify (using System Verilog) the expected behavior of that instruction. Any instruction operand values are defined symbolically, allowing the BMC tool to evaluate the properties for all possible operand values. Each such property is verified independently (i.e., when the pipeline doesn't contain other instructions). Hence, it is different from OCS-FV (which requires extensive manual work to avoid false failures as mentioned in Question 3.C). Properties similar to Single-I are used in industry, and automated approaches such as [Reid 16] may be used for generating such properties.

### 6. How long did it take to set up & run Symbolic QED?

*Setup for Design A*

Symbolic QED setup (using EDDI-V and the QED module in [Ganesan 18]) took 4 person-weeks (referred to as Symbolic QED: EDDI-V in Fig. 7). The primary task involved creating the QED module (which required understanding the design and the ISA including instruction encoding and register mapping, as discussed in Appendix A.2).

Enhanced EDDI-V for control flow errors and duplication using memory required 1 additional person-week each. This includes creating a memory model [Ecker 04], which avoids dramatic state space increase during BMC.

Single-I properties for *Design A* took 2 person-weeks. While we used a manual approach, automated techniques (Question 5.C) may be used.

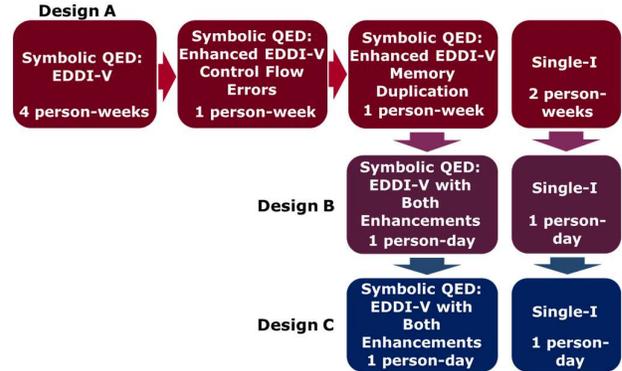

Fig. 7. Effort required for Symbolic QED setup.

*Setup for subsequent Designs B and C*

Adapting the EDDI-V setup (including EDDI-V enhancements) from *Design A* to *Designs B* and *C* required one person-day each. There were two significant design changes between *Design A* vs. *Designs B* and *C*: single-ROM interface in *B* and *C* (vs. dual-ROM interface in *A*), one additional instruction in *B* and *C* (vs. *A*). Adapting Single-I properties (manually) from *Design A* to *Designs B* and *C* took 1 person-day each.

Between design versions, it took practically no effort to adapt Symbolic QED (1 person-hour or less).

As shown in Table 1, Symbolic QED setup for the initial design took significantly less effort (8 person-weeks) vs. OCS-FV and CRS combined (17 person-months in the best case), i.e., by a factor of more than 8X. The benefits are even bigger for subsequent designs: 60X (2 person-days using Symbolic QED vs. 4-7 person-months using OCS-FV and CRS). Thus, Symbolic QED enables drastic improvement (i.e., reduction) in verification effort. Since DST efforts in Table 1 include design efforts as well and bugs detected by DST (and immediately fixed by designers) aren't reported, we conservatively attribute all DST efforts to design. In reality, Symbolic QED can potentially reduce DST efforts as well (although we cannot quantify that using our study).

Table 1: Setup effort: industrial verification flow vs. Symbolic QED.

|  | Total Setup Effort: Initial Design | Total Setup Effort: Subsequent Designs |
|---|---|---|
| DST* | 10 person-months | 1-3 person-months |
| OCS-FV | 5 person-months | 1 person-month |
| CRS | 12 person-months | 3-6 person-months |
| Total industrial verification flow (DST,OCS-FV,CRS) | 27 person-months (including DST) 17 person-months (excluding DST) | 5-10 person-months (including DST) 4-7 person-months (excluding DST) |
| Symbolic QED | 8 person-weeks | 2 person-days |

* – includes design effort as well

*Runtime*

As shown in Table 2 (BMC tool Onespin®, version 2018.2.1, from Onespin Solutions, running on an Intel(R) Xeon(R) E5-2690 v3 @ 2.6GHz with 32GB RAM), Symbolic QED with Enhanced EDDI-V (i.e., Enhanced EDDI-V targeting control flow errors as well as duplication using memory) took between 6 to 12 seconds to generate counterexamples. Single-I properties were even faster, finding counterexamples in 6 to 8 seconds.

Table 2: Bug Detection Runtime for Symbolic QED.

|  | Bug Detection Runtime (sec.) [min., avg., max.] |
|---|---|
| Symbolic QED with both EDDI-V Enhancements | [6, 8.3, 12] |
| Single-I | [6, 7.3, 8] |

### 7. What bugs did Symbolic QED detect?

*Symbolic QED detected all logic bugs (Fig. 8) detected by the industrial verification flow for all the 16 designs*. These bugs were related to both the RTL design as well as the design specification. The BMC tool generated Symbolic QED counterexamples starting from a state where the core is in the normal execution mode, the pipeline is empty, and all registers and memory location set to 0 (a *QED-consistent architectural state*, as explained in Question 5). The additional 7% in Fig. 8, detected uniquely by Symbolic QED, is related to specification for *Design A.v[f]*. Upon review with designers, we suspect that this 7% was perhaps also detected using the industrial verification flow (but wasn't reported and the specification document wasn't updated as well).

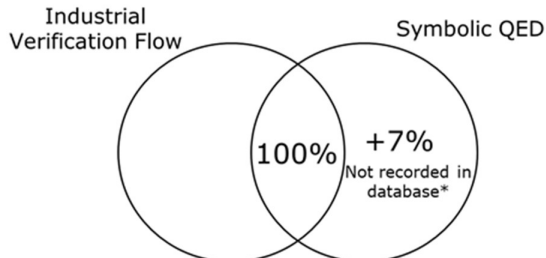

Fig. 8. Bugs detected by Symbolic QED (Enhanced EDDI-V: Control flow Errors; Enhanced EDDI-V: Duplication using Memory; Single-I) vs. industrial verification flow.

All these bugs were detected by only CRS in the industrial verification flow (Fig. 9); they escaped DST and OCS-FV. (Recall that bugs detected by DST weren't recorded, as discussed in Question 3.A). Upon detection, additional properties were generated manually to try to confirm these bugs using BMC; i.e., a specific property was generated for each bug *a posteriori* (after the bug became known). This further demonstrates the challenge of creating "good" properties for traditional BMC.

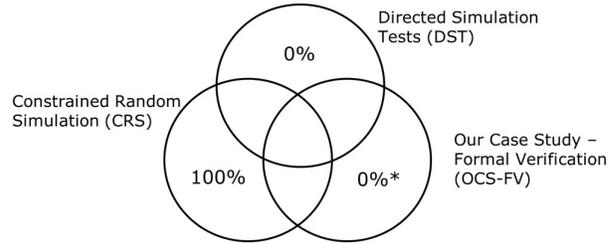

*One would expect OCS-FV to detect bugs detected by Single-I (Question 8, Fig. 10). That didn't happen in this study due to human error (explained in Question 8).

Fig. 9. Bugs detected by the industrial verification flow.

### 8. What bugs did various Symbolic QED features detect?

In Fig. 10, we show a breakdown of bugs detected by various Symbolic QED features discussed in Question 5. Symbolic QED using EDDI-V (Appendix A.1, with the updated QED module in [Ganesan 18]) detected 35.7% of the bugs. Enhanced EDDI-V (Question 5, A and B) uniquely detected another 35.7% of the bugs: 28.6% of these bugs caused control flow errors with wrong branch directions (detected by Enhanced EDDI-V using the QED-CF module) and the remaining 7.1% used instructions with a specific destination register (detected by Enhanced EDDI-V using memory for duplication).

The remaining 28.6% of bugs were detected by Single-I. One would expect the bugs detected by Single-I to be also detected by OCS-FV. However, due to human error (stemming from ways to cope with false failures as explained in Question 3.C), the OCS-FV properties missed certain details. As a result, the Single-I bugs were not detected by OCS-FV (until after bug detection by CRS when the properties were manually updated based on the detected bugs, as discussed in Question 7).

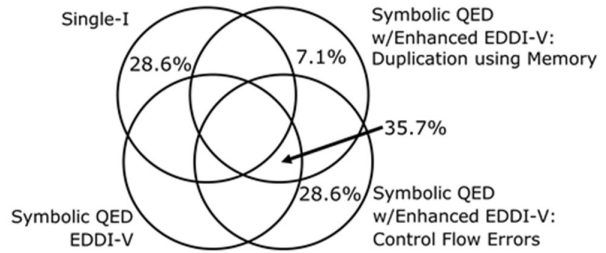

Fig. 10. Bugs detected by various Symbolic QED features.

### 9. How long did it take to debug using Symbolic QED?

The counterexamples generated by Symbolic QED were very short (Table 3). As a result, the debug time was similar to that of the industrial verification flow (less than a day to analyze and fix each bug).

Table 3: Counterexample length for Symbolic QED.

|  | Counterexample Length | |
|---|---|---|
|  | (cycles) [min., avg., max.] | (instructions) [min., avg., max.] |
| Symbolic QED with both EDDI-V Enhancements | [5, 7.4, 11] | [4, 6.2, 10] |
| Single Instruction | [2, 2, 2] | [1, 1, 1] |

### 10. What are some of the Symbolic QED future directions?

*More case studies:* While this study uses microcontroller cores, case studies on more complex (industrial or open-source) SoCs with processor cores, accelerators and uncore components can demonstrate further benefits of Symbolic QED.

*Symbolic QED with symbolic starting state:* Symbolic QED used in this case study uses a fixed starting state (Question 5, Appendix A.1). For bugs with very long activation sequences, the BMC tool may struggle to unroll the design far enough. Symbolic QED with symbolic starting states [Fadiheh 18, Ganesan 18] can overcome this challenge.

*Hardware security:* Techniques influenced by Symbolic QED can be highly effective in detecting security vulnerabilities in digital systems. Examples include: (a) the Unique Program Execution (*UPEC*) [Fadiheh 19] for detecting timing side channel-based security vulnerabilities in processors (an active research area, especially after the publication of Spectre and Meltdown attacks [Kocker 18, Lipp 18]); and, (b) hardware Trojan detection during pre-silicon verification using Symbolic QED with symbolic starting states [Ganesan 18].

*AMS-QED:* Symbolic QED techniques may be further extended to include analog-mixed-signal (*AMS*) design blocks, as explained in [Gielen 19].

*Bug localization during post-silicon validation and system operation:* Symbolic QED and E-QED techniques show promising results in localizing logic and electrical bugs during post-silicon validation and system operation [Singh 17, 18].

ACKNOWLEDGMENT

We acknowledge DARPA and the Stanford SystemX Alliance.

APPENDIX

### A.1. Symbolic QED

Symbolic QED searches all instructions sequences (within the BMC bound) and checks whether a QED test [Lin 14] exists that could fail. QED tests are based on software transformations that systematically ensure that errors created by a bug are detected quickly. The property analyzed by BMC is derived from the check that would detect the error during a QED test. Here, we focus on the EDDI-V (Error Detection using Duplicated Instructions for Validation) transformation. EDDI-V divides the registers (and memory) into two halves and creates consistent pairs between them, e.g. R0 and R16, R1 and R17, etc. It also duplicates instructions, creating two sequences (an *original* and a *duplicate* instruction sequence) that execute independently on each half of the data register and memory space. Therefore, an EDDI-V QED test starting from a *QED-consistent architectural state* (where all corresponding register and memory pairs are equal to each other, with no other instructions left to commit) will then execute the same instructions on each half of the registers and memory and, in the absence of a bug, result in a QED consistent state once the original and duplicate instruction sequences have executed. An inconsistency (different values in a register or memory pair) would indicate the presence of a bug. As a result, the BMC tool attempts to find a counterexample to the following property [Lin 15, Singh 18]:

$$qed\_ready \rightarrow \bigwedge_{a \in \{0..\frac{n}{2}-1\}} Ra == Ra',$$

where *n* is the number of registers defined by the ISA and *qed_ready* is asserted after the duplicate sequence is done. Here, (for a ∈ {0..n/2-1}), *Ra* and *Ra'* correspond to registers allocated for original instructions and duplicated instructions, respectively. Other QED transformations include *CFCSS-V* (Control Flow Checking using Software Signatures for Validation) targeting control flow instructions and *CFTSS-V* (Control Flow Tracking using Software Signatures for Validation) for detecting deadlocks, detailed in [Singh 18].

### A.2. QED Module

If a BMC tool does not impose any constraints on its inputs while analyzing a design for a counterexample, it may find false failures. To address this, environmental constraints (constraints that disallow invalid inputs) must be added, often through extensive manual effort. In Symbolic QED, the environmental constraints are straightforward. Specifically, starting from a *QED-consistent* state, the BMC tool must only consider sequences of instructions that correspond to QED tests while searching for a counterexample. Symbolic QED accomplishes this without extensive manual work using two mechanisms:

1. It constrains the inputs to the instruction fetch unit (of each core) to be arbitrary but valid instructions, directly obtained from the ISA.

2. It adds a *QED module* to the fetch unit of a processor core during BMC. The QED module automatically performs the EDDI-V transformation on-the-fly for any input sequence explored during BMC. The QED module is only used within the BMC tool and is not added to the manufactured IC; i.e., there is no performance/area/power overhead. The QED module is quite simple and needs to be designed only once for a given ISA, as shown in Question 7.

Creating the QED module requires understanding of the ISA encoding to constrain valid instructions and configure the module to transform original instructions into their duplicates. This also requires knowledge of the register mapping to ensure original and duplicate registers are correctly paired.

The QED module of [Lin 15, Singh 18] requires that all original instructions complete, followed by a waiting period for the pipeline to be flushed, and then duplicate instructions execute, before the qed_ready signal is asserted. The new QED module, used in this case study, from [Ganesan 18], instead allows arbitrary interleaving of the original and duplicate instruction subsequences. The QED module records the instructions generated by the BMC tool in a queue, adding to the queue each next original instruction and removing the instruction when its duplicate is executed.